\documentclass[aps,english,twocolumn,superscriptaddress,showkeys,noshowpacs]{revtex4}
\usepackage[T1]{fontenc}
\usepackage[utf8]{inputenc}
\usepackage{graphicx}
\usepackage{amssymb}
\usepackage{amsmath}
\usepackage{siunitx}

\newcommand{\D}{\mathrm{d}}

\newcommand{\bbm}{\begin{pmatrix}}
\newcommand{\ebm}{\end{pmatrix}}
\newcommand{\bma}{\begin{matrix}}
\newcommand{\ema}{\end{matrix}}
\newcommand{\bsm}{\begin{smallmatrix}}
\newcommand{\esm}{\end{smallmatrix}}
\newcommand{\bsbm}{\left( \begin{smallmatrix}}
\newcommand{\esbm}{\end{smallmatrix}\right)}

\newcommand{\vb}[1]{\left( #1 \right)}
\newcommand{\vsb}[1]{\left[ #1 \right]}

\newcommand{\mbf}[1]{\mathbf{#1}}

\newcommand{\dt}{\,\text{.}}

\newcommand{\usi}[2]{\ensuremath{#1 \, \si{#2}}}

\newcommand{\com}{\quad\text{,}}		
\newcommand{\eqb}{\nonumber\\ &\quad}

\allowdisplaybreaks[2]
\usepackage{babel}

\usepackage[normalem]{ulem} 
\usepackage{xcolor}

\definecolor{mgreen}{RGB}{21,105,26}

\begin{document}
\title{Metastable order protected by destructive many-body interference}

\author{M. Nuske}
\thanks{These authors have contributed equally to this work.}
\affiliation{Institut f\"ur Laserphysik, Universit\"at Hamburg, 22761 Hamburg, Germany}
\affiliation{Zentrum f\"ur Optische Quantentechnologien, Universit\"at Hamburg, 22761 Hamburg, Germany}
\affiliation{The Hamburg Center for Ultrafast Imaging, Luruper Chaussee 149, Hamburg 22761, Germany}
\author{J. Vargas}
\thanks{These authors have contributed equally to this work.}
\affiliation{Institut f\"ur Laserphysik, Universit\"at Hamburg, 22761 Hamburg, Germany}
\author{M. Hachmann}
\affiliation{Institut f\"ur Laserphysik, Universit\"at Hamburg, 22761 Hamburg, Germany}
\affiliation{Zentrum f\"ur Optische Quantentechnologien, Universit\"at Hamburg, 22761 Hamburg, Germany}
\author{R. Eichberger}
\affiliation{Institut f\"ur Laserphysik, Universit\"at Hamburg, 22761 Hamburg, Germany}
\affiliation{Zentrum f\"ur Optische Quantentechnologien, Universit\"at Hamburg, 22761 Hamburg, Germany}
\author{L. Mathey}
\affiliation{Institut f\"ur Laserphysik, Universit\"at Hamburg, 22761 Hamburg, Germany}
\affiliation{Zentrum f\"ur Optische Quantentechnologien, Universit\"at Hamburg, 22761 Hamburg, Germany}
\affiliation{The Hamburg Center for Ultrafast Imaging, Luruper Chaussee 149, Hamburg 22761, Germany}
\author{A. Hemmerich}
\affiliation{Institut f\"ur Laserphysik, Universit\"at Hamburg, 22761 Hamburg, Germany}
\affiliation{Zentrum f\"ur Optische Quantentechnologien, Universit\"at Hamburg, 22761 Hamburg, Germany}
\affiliation{The Hamburg Center for Ultrafast Imaging, Luruper Chaussee 149, Hamburg 22761, Germany}

\begin{abstract}
The phenomenon of metastability can shape dynamical processes on all temporal and spatial scales. Here, we induce metastable dynamics by pumping ultracold bosonic atoms from the lowest band of an optical lattice to an excitation band, via a sudden quench of the unit cell. The subsequent relaxation process to the lowest band displays a sequence of stages, which include a metastable stage, during which the atom loss from the excitation band is strongly suppressed. Using classical-field simulations and analytical arguments, we provide an explanation for this experimental observation, in which we show that the transient condensed state of the atoms in the excitation band is a dark state with regard to collisional decay and tunneling to a low-energy orbital. Therefore the metastable state is stabilized by destructive interference due the chiral phase pattern of the condensed state. Our experimental and theoretical study provides a detailed understanding of the different stages of a paradigmatic example of many-body relaxation dynamics.
\end{abstract}

\keywords{Atomic and Molecular Physics, Condensed Matter Physics, Quantum Physics}

\pacs{03.65.Yz, 03.75.Lm, 03.75.Kk, 05.65.+b, 05.70.Ln} 

\maketitle
The relaxation dynamics of a many-body system that has been driven out of equilibrium can either be governed by a single time scale, as in an exponential decay process, or take a more intricate form \cite{Fer:12}. In the latter case, the relaxation dynamics of the system might first progress towards a long-lived, metastable state, before relaxing to thermal equilibrium \cite{Mor:18,Mal:19}. The time evolution of this scenario naturally separates into three stages. The first stage is the relaxation to the metastable state. The second one is the long-lived metastable state itself, the third stage is the relaxation from the metastable state to equilibrium. A common origin of the long lifetime of the metastable state is the existence of a free energy barrier that inhibits relaxation at sufficiently low temperatures \cite{And:72, Bra:06, Kor:05, Sme:97, Kua:00, Kas:01}. Furthermore, the phase coherent properties of a system evolving according to quantum dynamics can enhance or suppress transitions via constructive or destructive interference. A striking example for long-lived states that are stabilized by destructive interference are dark states in the electronic structure of atoms or molecules \cite{Alz:76}, which are at the basis of prominent phenomena as electromagnetic transparency \cite{Har:97}. For the example of a three-level system, the dark state emerges due to destructive interference of the coupling to the two other states. We note that the phase coherence of the quantum state is imperative for the stability of the dark state, as it cannot be understood within a rate equation description of the system. Finally, we note that metastable states of Bose-Einstein condensates, and their dynamics in general, are strongly modified due to bosonic enhancement and statistics, in comparison to a classical gas \cite{Mie:99}. Further studies on metastable dynamics have been reported in Refs.~\cite{Ber:17, Tur:18, Ho:19, Dro:19}.

\begin{figure}[tb]
\centering
\includegraphics[width=1\linewidth]{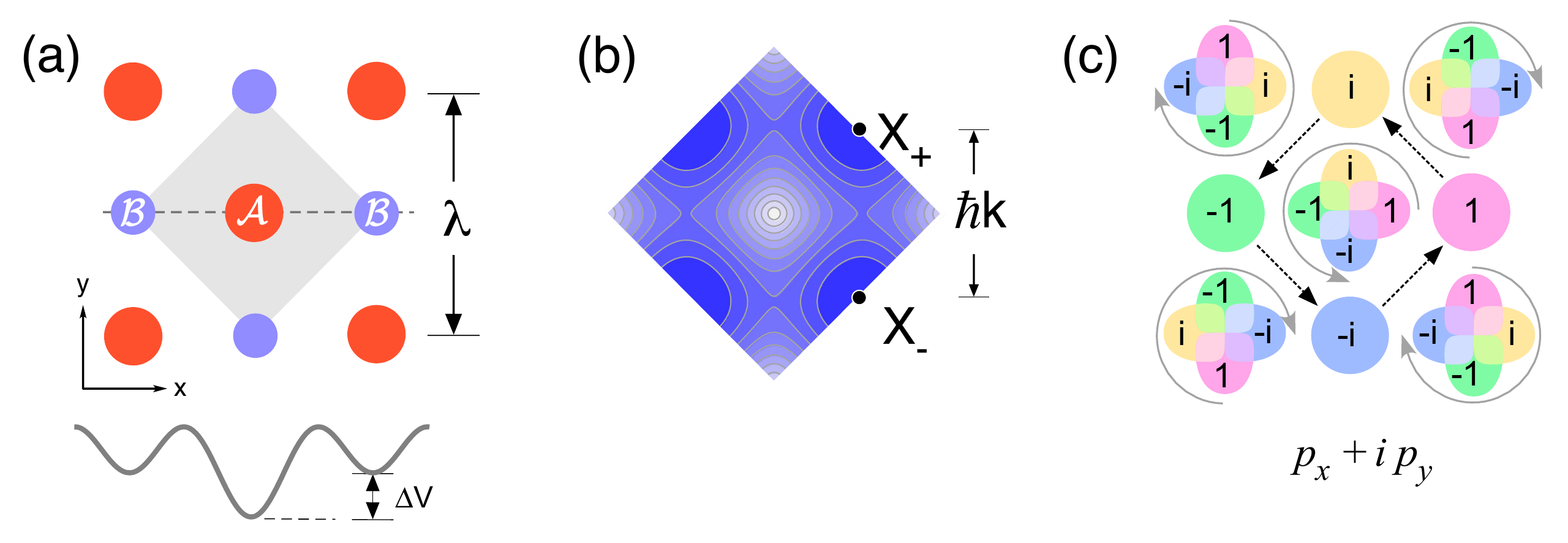}
\caption{(a) A single plaquette of the chequerboard lattice comprising $\mathcal{A}$ and $\mathcal{B}$-wells with tunable well depths difference $\Delta V$. Here, $\lambda$ is the lattice wavelength and $k = 2\pi/\lambda$ the wave vector. (b) The second Bloch band of the lattice in (a), plotted across the first Brillouin zone with the two inequivalent energy minima at $X_{\pm}$ highlighted. Blue denotes low and white denotes high energy. (c) Sketch of the many-body ground state wave function of the second band. The deep $\mathcal{A}$-wells host $p_x$- and $p_y$-orbitals, depicted by dumbbells. These are phase locked by interaction to form $(p_x \pm i\,p_y)$-orbitals. The shallow $\mathcal{B}$ wells host $s$-orbitals. The locations, where the local phase $\phi$ takes the values $e^{i\phi} \in \{1,i,-1,-i\}$, are indicated by colors, showing a chiral pattern of staggered currents.}
\label{fig:lattice}
\end{figure}

\begin{figure}[h!]
\centering
\includegraphics[width=0.9\linewidth]{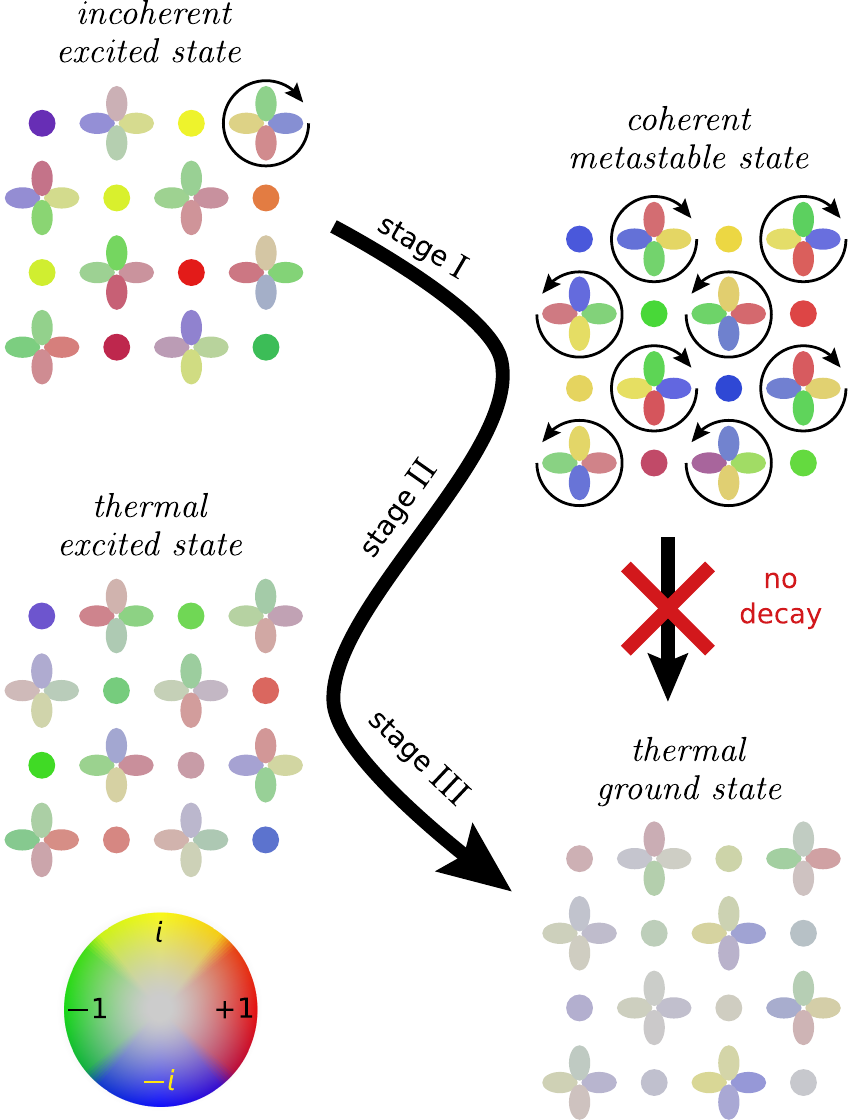}
\caption{The time evolution passes through four states connected by three stages of relaxation I, II and III, as explained in the text. All states are plotted in terms of the occupations of local $s$-orbitals, depicted by disks, and $p$-orbitals, depicted by dumbbells, obtained for a snapshot of a single $xy$-layer of our numerical simulation. The colours indicate the local phases according to the color wheel in the left lower corner and amplitudes are indicated by contrast, where light gray indicates low amplitude. For the sake of clarity, we show the idealized case of very low initial temperature, $T=\usi{0.5}{\nano\kelvin}$, that leads to nearly perfect phase coherence after stage I.}
\label{fig:decayPhases}
\end{figure}

Metastability is not only an important concept for our understanding of thermalization processes, but can also be used as a starting point to study states that are inaccessible in the equilibrium phase diagram. A well-defined platform to study a many-body system far out of equilibrium are ultracold atomic gases in general and in particular, when prepared in excited Bloch bands of an optical lattice \cite{Lew:11}. Exciting atoms to higher bands gives access to orbital physics, which also plays a crucial role for metal-insulator transitions, superconductivity and colossal magnetoresistance in transition-metal oxides \cite{Tok:00, Mae:04}. Ultracold atoms in excited bands are naturally prone to relaxation towards the lowest band, however, their orbital nature opens a promising arena for studying scenarios of metastability.

In this work we present an experimental and theoretical study of metastable dynamics of ultracold atoms in the second band of an optical lattice. We provide a theoretical interpretation of the observation of inhibited decay based on numerical and analytical arguments that identify destructive quantum interference of distinct decay channels as the underlying mechanism.

\begin{figure*}[htb]
\centering
\includegraphics[width=\linewidth]{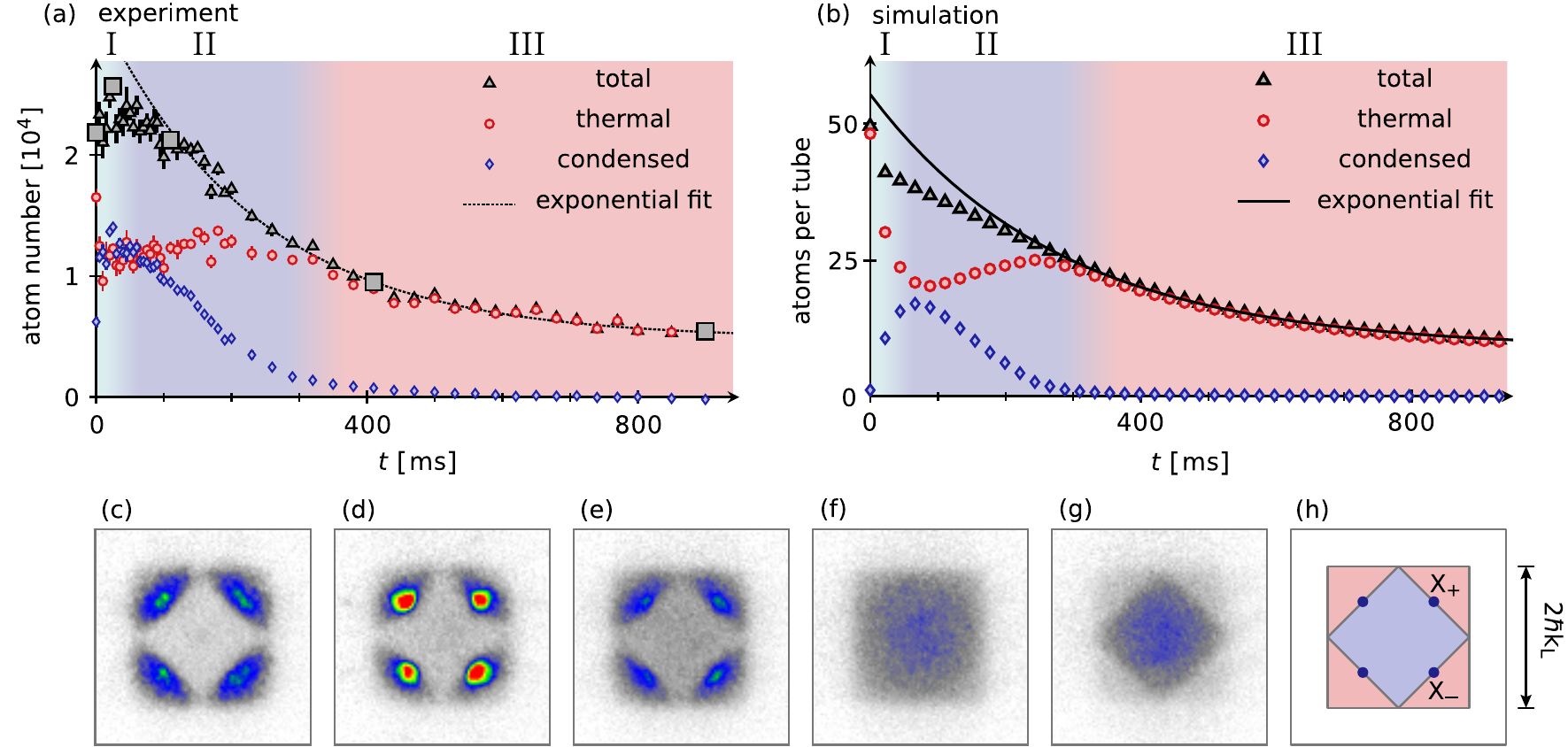}
\caption{Time evolution of the total number of atoms, the condensate and the thermal fraction in the second band after a quench to $\Delta V=\usi{0.43}{V_0}$. Panel (a) shows experimental and panel (b) numerical data. The different background colours identify the three relaxation stages of Fig.~\ref{fig:decayPhases}. Black lines show exponential fits to the data points in stage III. (c-g) Time of flight images for experimental data points marked by black squares in (a). We perform band mapping, such that the resulting time of flight image depicts the population of quasi-momentum space. Panel (h) illustrates the first and second Brillouin zone and the $X_{\pm}$-points of the lattice. The temperature of the initial state for both experiments and simulation is $T\approx\usi{0.5}{E_{\rm rec}/k_B}\approx\usi{50}{\nano\kelvin}$. We note that atom numbers for experiment and simulation are adjusted to be comparable in the center of each tube as is described in Appendix \ref{app:classicalFieldTheory}.}
\label{fig:selfStabilization}
\end{figure*}

Our experiments begin with the preparation of an ultracold gas of bosonic atoms in the ground state of an optical lattice. This lattice is composed of shallow and deep potential wells, forming a checkerboard pattern in the $xy$-plane, see Fig.~\ref{fig:lattice}(a). The atoms are weakly bound along the $z$-direction by a harmonic trap potential, such that the lattice wells acquire a tubular shape. The atoms are then pumped into the first excitation band by a sudden quench of the unit cell. This induces relaxation dynamics to the lowest band that we observe in the momentum-resolved band occupation, obtained via band mapping, see Figs.~\ref{fig:selfStabilization}(c--g). We thus derive the total atom number, and the condensed atom number, see Fig.~\ref{fig:selfStabilization}(a). Initially, the atoms are found to condense in the excitation band. Subsequently the system forms a metastable state in which the total number of atoms changes only slowly, while the number of condensed atoms decreases. When the condensed fraction reaches zero, the total atom number decays exponentially.

We present a detailed theoretical understanding of the relaxation dynamics. We identify three relaxation stages illustrated in Fig.~\ref{fig:decayPhases}, in which we depict snapshots of our semi-classical c-field simulation, see App.~\ref{app:classicalFieldTheory} and Ref.~\cite{Sin:15,Sin:20}. Directly after the quench, the atomic gas forms an \textit{incoherent excited state}. In stage I the atoms condense, the phase coherence increases and the atoms form a \textit{coherent metastable state}, which we identify as the metastable state observed in experiment. This condensate has been predicted and detected in previous work \cite{Liu:06, Sto:08, Li:16,Wir:11, Oel:11, Oel:12, Oel:13, DiL:14, Koc:15, Koc:16}. It has a finite occupation of $p$-orbitals in the deep wells and of $s$-orbitals in the shallow wells, see Fig.~\ref{fig:lattice}(c). We confirm the proposed chiral phase winding on $p$-orbitals within our simulations, see Fig.~\ref{fig:decayPhases}: \textit{coherent metastable state}. The next stage, stage II, is characterized by a comparatively slow decay and a finite amount of condensed atoms. At the onset of stage III, the \textit{coherent metastable state} is entirely converted to the \textit{thermal excited state}, which relaxes exponentially towards the \textit{thermal ground state}. 

By combining the results of our numerical simulations with analytical arguments we identify the origin of the observed metastability as destructive interference. This interference effect arises due to the chiral phase texture of the condensate. The chiral phase winding on $p$-orbitals induces a staggered order of the $s$-orbitals, see Fig.~\ref{fig:lattice}(c). Therefore the wave function of $s$-orbital atoms on opposing sides of a given deep well has opposite sign and the corresponding hopping processes interfere destructively. We find a similar mechanism of destructive interference for the second decay channel, which is an interaction-induced decay where two $p$-orbital atoms collide and scatter into the $s$-orbital on the same site. As a consequence of these interference mechanisms, the condensate with respect to collisions constitutes a many-body dark state and therefore direct relaxation of the \textit{coherent metastable state} to the \textit{thermal ground state} is inhibited, see Fig.~\ref{fig:decayPhases}. Instead, decay arises solely via the {\it thermal excited state}.

\subsection*{EXPERIMENTAL AND THEORETICAL SETUP}
\begin{figure}[h]
\centering
\includegraphics[width=0.9\linewidth]{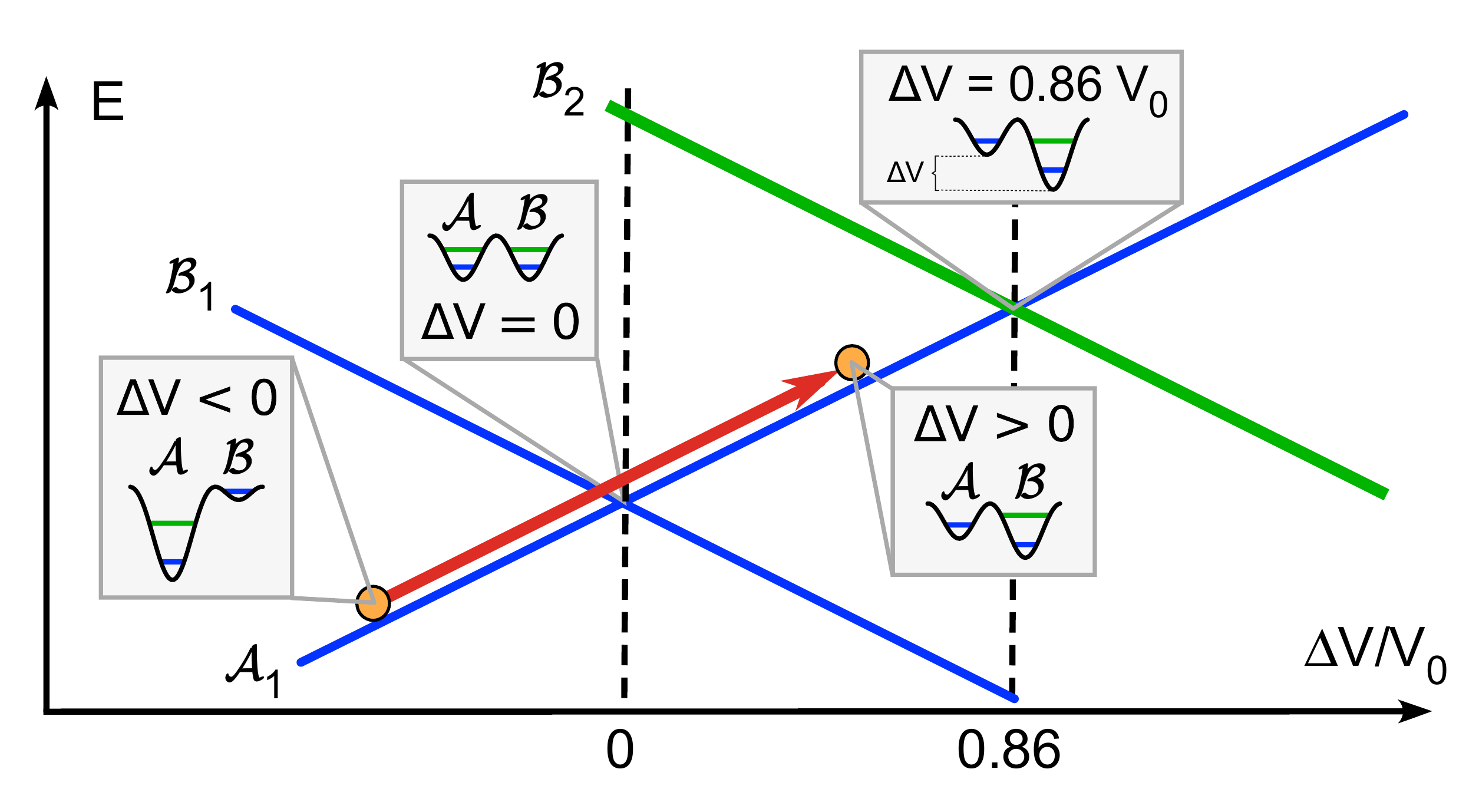}
\caption{The atoms are prepared in the lowest band at large negative offset $\Delta V<0$. The red arrow depicts a sudden quench of $300\,{\rm \mu s}$ duration to $\Delta V>0$ that transfers the atoms into the second band. Insets show exemplary $\mathcal{A}$ and $\mathcal{B}$-wells with horizontal lines indicating $s$- and $p$-orbital energies. Ascending and descending solid lines show the energies at zero quasi-momentum of the lowest band of the $\mathcal{A}$ sublattice and the three lowest bands of the $\mathcal{B}$ sublattice, respectively.}
\label{fig:sketchLoadingMech}
\end{figure}

A BEC of $10^5$ $^{87}{\rm Rb}$ atoms is prepared in the $|{\rm F}=2, {\rm m_{F}}=2\rangle$ hyperfine state in an isotropic magnetic trap. The BEC is adiabatically loaded into the lowest band of the double-well chequerboard lattice with the two inequivalent sublattice sites denoted $\mathcal{A}$ and $\mathcal{B}$. The lattice is formed by laser beams with a wavelength $\lambda = 1064\,$nm, see Fig.~\ref{fig:lattice}(a). For more details see Appendix~\ref{app:Exp} and also Refs.~\cite{Wir:11, Oel:11, Oel:12, Oel:13, DiL:14, Koc:15, Koc:16}. The key tool is the tunability of the potential offset $\Delta V$ between the $\mathcal{A}$ and $\mathcal{B}$ sublattice wells, see Fig.~\ref{fig:sketchLoadingMech}. Initially, the $\mathcal{A}$-site wells are deep and the $\mathcal{B}$-site wells are shallow such that $\Delta V$ is negative. We load the BEC at large negative offset $\Delta V\ll 0$, such that the resulting ground state has vanishing occupation on the $\mathcal{B}$-sublattice sites. The atoms, which exclusively reside on the $\mathcal{A}$-sublattice, are phase-incoherent even for temperatures as low as \usi{50}{\nano\kelvin} due to on-site particle number squeezing resulting from the sizable repulsive interaction in the deep lattice wells in the $xy$-plane. We perform a rapid quench within $300\,{\rm \mu s}$ to $\Delta V>0$ such that the two lowest bands cross and a significant fraction of atoms is excited to the second band (cf. Fig.~\ref{fig:sketchLoadingMech}). In order to study the resulting relaxation dynamics the atoms are observed via a band mapping technique. We extract the condensed and thermal fractions in the second band as explained in Appendix~\ref{app:Exp}. 

For our c-field simulations we use a tight-binding model for the four lowest bands of the lattice and solve the time evolution by computing classical equations of motion for an ensemble of complex-valued fields. The details of this algorithm are described in Appendix~\ref{app:classicalFieldTheory} and Refs.~\cite{Bla:08, Gar:01, Gar:03, Coc:09, Pol:10, Dav:11}. We model both the dynamics within the lattice in the $xy$-plane as well as the dynamics within the tubes in the $z$-direction. The same quench of $\Delta V$ as in the experimental procedure excites the atoms into the second band. In order to model the band mapping process we project the wave function onto the Bloch functions of the tight-binding lattice. 

\subsection*{RELAXATION DYNAMICS}
The central experimental observation is the momentum resolved occupation of the bands, obtained via band mapping, and depicted in Fig.~\ref{fig:selfStabilization}(c--g). From these measurements we derive the number of condensed and thermal atoms, which evolve in time as shown in Fig.~\ref{fig:selfStabilization}(a).
We simulate a similar protocol numerically, and depict the resulting number of condensed and thermal atoms per tube in Fig.~\ref{fig:selfStabilization}(b).

Immediately after the quench, the condensate fraction vanishes and all available Bloch modes in the second band are nearly equally populated, see Fig.~\ref{fig:selfStabilization}(c). The relaxation dynamics begins with a condensation stage, denoted stage I, where the number of condensed atoms increases and the number of thermal atoms decreases. In the experimental data in Fig.~\ref{fig:selfStabilization}(a) this condensation stage takes 20~ms, while in the numerical simulations in Fig.~\ref{fig:selfStabilization}(b) it occurs on a slower time scale of 70~ms. We explain in Appendix \ref{app:compareTubeLength} that the increased duration is a result of using shorter tubes in the $z$-direction for our numerical simulations. The formation of coherence within the lattice plane ($xy$) is associated with increased phase fluctuations along the $z$-direction \cite{Pau:13}. Despite the different durations of the condensation stage, qualitatively, the same dynamical features are found in the observations and calculations. We find that the role of the tubes in the $z$-direction as an entropy reservoir is crucial for the initial condensation stage. In fact, the atoms would not condense for an effective 2-dimensional system, see Appendix \ref{app:compareTubeLength}. 

At the beginning of stage II there is a large number of condensed atoms, which manifests itself in the accumulation of a significant fraction of atoms in the two inequivalent minima of the second band, i.e., see Fig.~\ref{fig:selfStabilization}(d). These minima are the $X_{\pm}$-points as depicted in Fig.~\ref{fig:lattice}(b). As has been shown in previous work, the atoms form a condensate with a chiral phase texture composed of vortical currents with opposite signs for adjacent plaquettes \cite{Koc:15}. The numerical data depicted in the second panel of Fig.~\ref{fig:decayPhases} for the {\it coherent metastable state}, suggests that this order emerges dynamically as a result of the recondensation process.

Stage II is characterized by a finite number of condensed atoms, a stable or even increasing number of thermal atoms and a slower than exponential decay of the total number of atoms. The last observation can be seen from the deviation of the total number of atoms from an exponential fit in Figs.~\ref{fig:selfStabilization}(a) and (b). In the numerical simulation the decay is almost linear during stage II. At the beginning of stage III, the critical temperature for condensation is reached and only thermal atoms remain, see Fig.~\ref{fig:selfStabilization}(f). At this point, the atoms decay exponentially as is predicted in Fig.~\ref{fig:selfStabilization}(b) and confirmed by the observations in (a). Eventually at the end of stage III all atoms have relaxed to the lower band with only small thermal excitations of the upper band, see Fig.~\ref{fig:selfStabilization}(g).

We identify the initial slow decay during stage II as a signature of the proposed mechanism for metastability. As mentioned above and as we will describe in more detail in the following section, the chiral phase pattern, as observed in our simulations, stabilizes the condensate against decay to the first Bloch band through destructive interference of the dominant decay channels. In fact, the condensate itself is expected to show perfectly balanced destructive interference and hence represents a dark state with infinite lifetime. Decay should arise exclusively via the thermal fraction of atoms in the second band, see {\it thermal excited state} in Fig.~\ref{fig:decayPhases}. In other words, condensate atoms become thermal atoms, which can decay to the lowest band. 
This prediction determines the dynamics during stage II. The underlying physics leading to this behavior is as follows: the thermal fraction of the second band relaxes to the lowest band, whereby significant kinetic energy is gained, which heats the thermal atoms in the second band through collisions with decayed atoms. This in turn depletes the condensate fraction into the thermal fraction and hence overcompensates its exponential decay. 
The remarkable agreement with the observations in Fig.~\ref{fig:selfStabilization}(a) shows that direct relaxation of the condensate towards the lowest band is inhibited, as predicted by the model in (b). Calculations and corresponding observations deferred to App.~\ref{app:thermalDecay} also show that a higher initial temperature with an initially larger thermal fraction leads to an earlier onset of stage III and hence a significantly faster decay.

\begin{figure}[tb]
\centering
\includegraphics[width=0.9\linewidth]{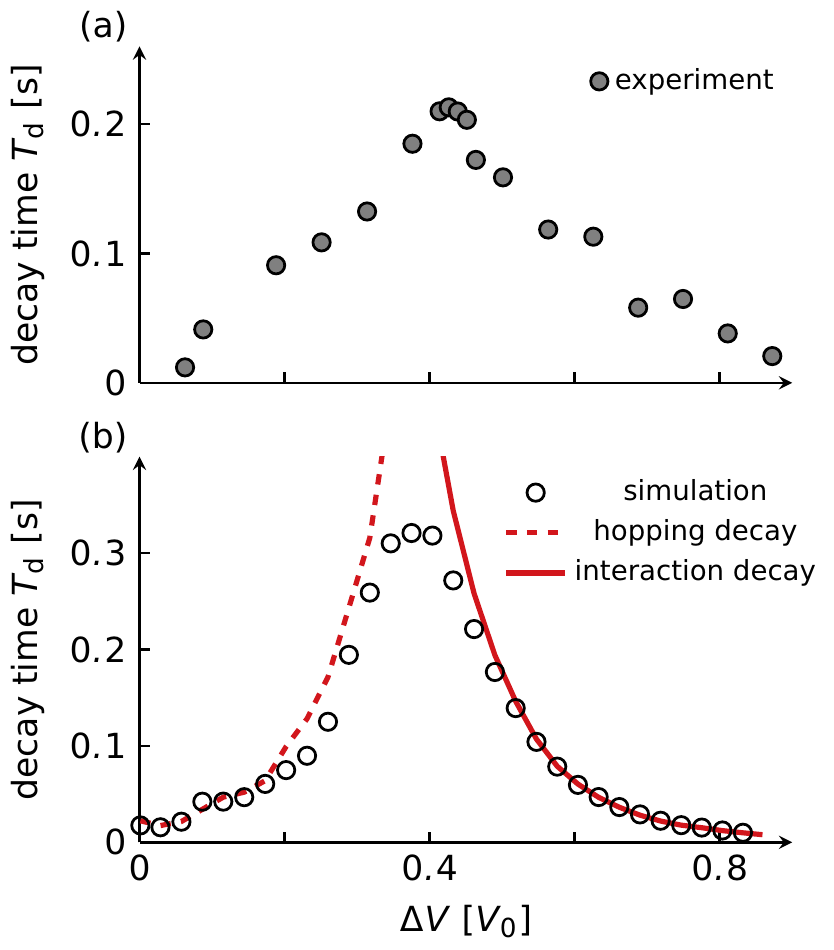}
\caption{Time scales of the exponential decay in stage III as a function of the final potential offset $\Delta V$ approached in the quench.
(a) and (b) show observations and numerical results, respectively. The dashed and solid red line graphs show calculations for $W_{\textrm{int}}$ and $W_{\textrm{hop}}$ set to zero, respectively. The temperature of the initial state for (a) and (b) is $T\approx\usi{0.5}{E_{\rm rec}/k_B}\approx\usi{50}{\nano\kelvin}$.}
\label{fig:decayOfTheta}
\end{figure}

\subsection*{NATURE OF DECAY MECHANISM}
In order to understand why the chiral phase pattern of the condensate inhibits decay, we analyze the contributing decay mechanisms. The character of the dominant decay channels depends on the value of the potential offset $\Delta V$. We consider values of $\Delta V$ between the first two band crossings in Fig.~\ref{fig:sketchLoadingMech}. For $\Delta V=0$ the sublattice sites have equal depth and hence their $s$-orbitals are degenerate, while at the second band crossing for $\Delta V= \usi{0.86}{V_0}$ the $p$-orbitals on $\mathcal{B}$-sublattice sites are degenerate with $s$-orbitals on $\mathcal{A}$-sublattice sites. Since only the thermal fraction of atoms in the second band is expected to decay to the lowest band, we utilize the exponential decay during stage III, to determine the decay times observed for different final values of $\Delta V$ shown in Fig.~\ref{fig:decayOfTheta}(a), see Appendix~\ref{app:Exp}. The decay is slow for intermediate values of $\Delta V$, while becoming increasingly fast towards the two band crossings at $\Delta V=0$ and $\Delta V=\usi{0.86}{V_0}$. 

Our simulation leads to the white circles in Fig.~\ref{fig:decayOfTheta}(b), which shows good qualitative agreement with (a), particularly reproducing the life time maximum close to $\Delta V/V_0 = 0.4$. The simulation, however, provides additional insight into the character of the distinct underlying relaxation processes on both sides of that maximum. The main decay channel on the right side close to $\Delta V=\usi{0.86}{V_0}$ is an interaction-driven decay, as has been already proposed in Ref.~\cite{Pau:13}. In this regime the $p$-orbitals on $\mathcal{B}$-sites are nearly degenerate with $s$-orbitals on $\mathcal{A}$-sites and hence are strongly occupied. Therefore, the dominant interaction term leading to band relaxation is associated with the collision of two $p_x$- or $p_y$-atoms on a deep $\mathcal{B}$-well at site $\mbf R$, which are both scattered into the $s$-orbital on the same site:
\begin{align}
W_{\textrm{int}} \equiv \sum_{\mbf R \in \mathcal{B}}  b_{{\mbf R},s}^\dag b_{{\mbf R},s}^\dag \vb{b_{{\mbf R},x}b_{{\mbf R},x}+b_{{\mbf R},y} b_{{\mbf R},y}}   \dt
\end{align}
Here $b_{{\mbf R},\{x,y,s\}}^\dag$ ($b_{{\mbf R},\{x,y,s\}}$) creates (annihilates) an atom on site ${\mbf R}$ in orbital $\{p_x,p_y,s\}$. On the left side of the maximum, however, close to $\Delta V = 0$, we identify the main decay channel to be associated with an atom in the $s$-orbital of a shallow well, belonging to the second band, hopping to the $s$-orbital of an adjacent deep well, belonging to the first band. This process is expected to be increasingly strong the closer the involved $s$-orbitals are to degeneracy. The responsible decay term is
\begin{align}
W_{\textrm{hop}} \equiv  \sum_{\mbf R \in \mathcal{B}, \nu \in \{x,y\}} b_{\mbf R,s}^\dag  (b_{{\mbf R}-{\mbf e}_{\nu} ,s} + b_{{\mbf R}+{\mbf e}_{\nu} ,s}  ) \com 
\end{align}
where the vectors ${\mbf e}_{x}, {\mbf e}_{y}$ connect $\mathcal{A}$ and $\mathcal{B}$ sites along the $x$ and $y$ directions, respectively. We confirm these considerations by selectively switching off either of the terms $W_{\textrm{int}}$ and $W_{\textrm{hop}}$ for the entire duration of our simulation. If $W_{\textrm{int}}$ is set to zero, we find the dashed red line in Fig.~\ref{fig:decayOfTheta}, which only accounts for $W_{\textrm{hop}}$. As expected, the decay times given by the white disks are well approximated on the left side of the life time maximum. Similarly, for $W_{\textrm{hop}} = 0$, the red solid line is found in Fig.~\ref{fig:decayOfTheta}, which provides a decent approximation on the right side of the life time maximum. For intermediate values of $\Delta V$ there is a competition between the two decay channels. The maximal life time is observed close to $\Delta V=\usi{0.4}{V_0}$. Calculations predict a slight shift depending on e.g.~temperature, see App.~\ref{app:decayTimeScaleDiffT}.

\begin{figure}[tb]
\centering
\includegraphics[width=1\linewidth]{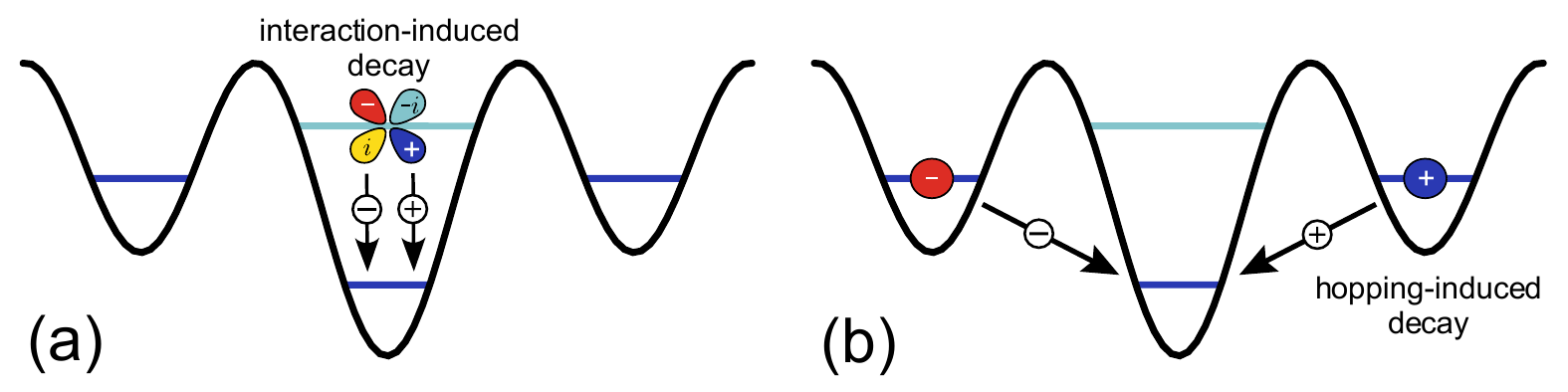}
\caption{Sketch of the two dominant decay mechanisms and the inhibition of decay for the condensate mode. (a) shows the case of interaction-induced decay, where two $p_x$ atoms or two $p_y$ atoms in a deep well collide and are transferred to the $s$-orbital in the same well. For condensate atoms populating $(p_x \pm i p_y)$-orbitals, the transition amplitudes for these two processes interfere destructively. (b) shows the case of hopping-induced decay: an atom in the $s$-orbital of a shallow well, belonging to the second band, hops to the $s$-orbital of an adjacent deep well, belonging to the first band. Condensate atoms in the $s$-orbitals of the two shallow wells on opposite sides of a deep well have opposite local phases. Hence the tunneling of such atoms to the $s$-orbital of the deep well destructively interferes.}
\label{fig:destrucInt}
\end{figure}

Next, we consider the effect of the two relaxation channels $W_{\textrm{int}}$ and $W_{\textrm{hop}}$ on the condensate fraction. This chiral condensate, denoted $|p_x \pm i p_y \rangle$, can be well approximated as a product of local coherent states, residing at each lattice site. More precisely, $|p_x \pm i p_y \rangle = |\mathcal{A} \rangle \otimes |\mathcal{B} \rangle$ with 
\begin{eqnarray}
\label{eq:Decay2b}
|\mathcal{A} \rangle &=& \prod_{\mbf R\in \mathcal{A}} |s(\mbf R)\rangle 
\\ \nonumber
|\mathcal{B} \rangle &=& \prod_{\mbf R\in \mathcal{B}} |p_x(\mbf R)\rangle \otimes |p_y(\mbf R)\rangle \otimes |0\rangle_ {\mbf R,s} \,.
\end{eqnarray}
Here, $|\mathcal{A} \rangle$ comprises local coherent states with on average $|s(\mbf R)|^2$ atoms in the $s$-orbitals of the shallow wells, while $|\mathcal{B} \rangle$ accounts for coherent states with on average $|p_x(\mbf R)|^2$ atoms in the $p_x$-orbitals and $|p_y(\mbf R)|^2$ atoms in the $p_y$-orbitals in the deep wells. Note also the empty $s$-orbitals in the deep wells, denoted $|0\rangle_ {\mbf R,s}$. The phases of the complex functions $s(\mbf R)$ on the $\mathcal{A}$-sublattice and $p_x(\mbf R), p_y(\mbf R)$ on the $\mathcal{B}$-sublattice take values in $\{1,i,-1,-i\}$, thus forming the chiral phase texture, shown in Fig.~\ref{fig:lattice}(c). This phase texture imposes the relation $p_y(\mbf R) = \pm i\, (-1)^{\nu+\mu} p_x(\mbf R)$ with $\mbf R = (\mu+\nu)\,  \mbf e_x + (\mu - \nu) \, \mbf  e_y$, with the interaction-induced relative phase factor $\pm i$ and a sign change $(-1)^{\nu+\mu}$ for adjacent $\mathcal{B}$-sites \cite{Oel:13}. As a consequence, using $b_{\mbf R,x} |p_x(\mbf R)\rangle = p_x(\mbf R) \,|p_x(\mbf R)\rangle$ and $b_{\mbf R,y} |p_y(\mbf R)\rangle = p_y(\mbf R) \,|p_y(\mbf R)\rangle$, and assuming equal populations of $p_x$- and $p_y$-orbitals, i.e. $|p_x(\mbf R)|^2 = |p_y(\mbf R)|^2$, one finds
\begin{align}
\label{eq:Decay2}
W_{\textrm{int}} |p_x \pm i p_y \rangle = 0 
\end{align}
and hence perfect destructive interference. This process is schematically illustrated in Fig.~\ref{fig:destrucInt}(a). The annihilation of two atoms in a $p_x$-orbital or in a $p_y$-orbital, phase-shifted by $\pi/2$, leads to different signs, such that both transition amplitudes eliminate each other.   

A consideration similar to that for $W_{\textrm{int}}$ shows that the condensate is also stable with respect to the hopping-induced decay channel, i.e., 
\begin{align}
W_{\textrm{hop}} |p_x \pm i p_y \rangle = 0\,. \end{align}
As illustrated in Fig.~\ref{fig:destrucInt}(b), condensate atoms in the $s$-orbitals of the two shallow wells on opposite sides of a deep well have opposite phases. Hence the tunneling of such atoms to the $s$-orbital of the deep well destructively interferes.	

\subsection*{TWO-FLUID MODEL}
Our understanding of the dynamics in stages I, II and III of Fig.~\ref{fig:selfStabilization} suggests the concept of two interacting fluids, i.e., a stable condensate in equilibrium with a thermal fraction that decays and thereby heats. In stage I, the thermal fraction exceeds the value compatible with the present temperature and hence a condensate fraction forms. In stage II, the condensate fraction is again redistributed to the thermal fraction, while it undergoes decay and heating. In stage III, the condensate has vanished and the thermal fraction decays exponentially. We capture this dynamical mechanism with a simplified two-fluid model which reproduces the dynamics in stages I, II and III and gives qualitative insight into the associated heating dynamics. The two-fluid model is not expected to deliver quantitative information on timescales or fractional populations, and hence is formulated in terms of unitless quantities. 

\begin{figure}[tb]
\centering
\includegraphics[width=0.9\linewidth]{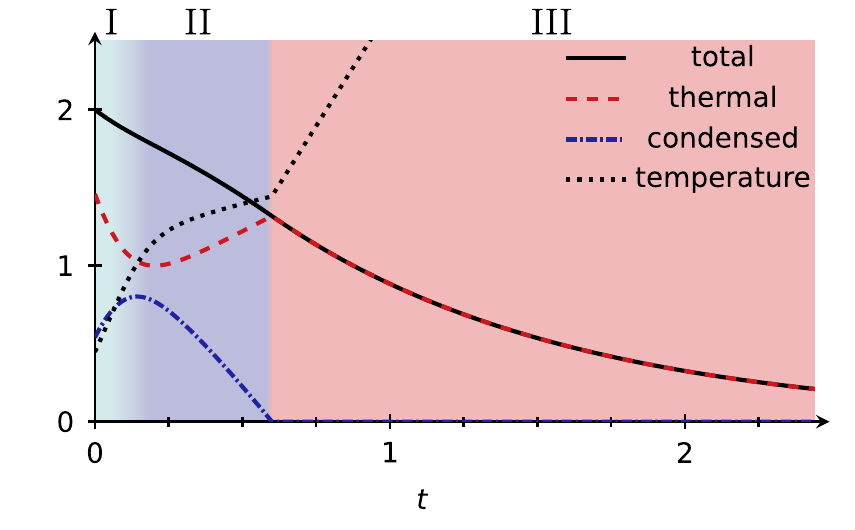}
\caption{Two-fluid model for self-stabilization. We show the total, thermal and condensed particle number as well as temperature computed within a two-fluid model, for details see App.~\ref{app:twoFluidModel}. The model shows qualitative agreement with experimental and simulated data shown in Fig.~\ref{fig:selfStabilization}. All quantities within the two-fluid model are without units. We use $\gamma_{\rm dec}=1$, $\Gamma_{\rm eq}=3$, $\alpha=2$ and $\Delta E=3$.}
\label{fig:twoFluidModelMain}
\end{figure}

We split the total number of atoms into a thermal and a condensed part $N_{\rm tot}=N_{\rm th}+N_{\rm c}$. The thermal part decays exponentially with constant $\gamma_{\rm dec}$, while the condensed part is stable. Each decaying atom gains an energy $\Delta E$ proportional to the band gap between the first excited and lowest band. This additional energy is assumed to thermalize among all atoms and hence increase the temperature $T$. Additionally, we assume that thermalization between the thermal and the condensed fraction occurs at fixed total energy on a time-scale $\Gamma_{\rm eq}$. The resulting differential equations are
\begin{align}
\label{eq:TFModel}
\dot N_{\rm th}&=-\gamma_{\rm dec} N_{\rm th} + \Gamma_{\rm eq} (T^\alpha-N_{\rm th}) \nonumber \\
\dot N_{\rm tot}&=-\gamma_{\rm dec} N_{\rm th} \\ \nonumber 
\dot T&=\gamma_{\rm dec} \Delta E - \frac{\Gamma_{\rm eq} T}{N_{\rm th}} (T^\alpha-N_{\rm th})\com
\end{align}
where $T$ is scaled in terms of an arbitrary unit temperature, an exponent $\alpha$ has been introduced and further details are described in App.~\ref{app:twoFluidModel}. Equation \ref{eq:TFModel} explicitly encodes the assumption that the condensed number of atoms does not decay, as the atom loss only depends on the thermal atom number.
We find that the resulting time evolution, shown in Fig.~\ref{fig:twoFluidModelMain}, captures the different decay stages and hence stresses the general applicability of the inhibited-decay mechanism presented in this paper. Furthermore, the validity of the two-fluid description is highlighted by an approximate analytical solution that predicts linear scaling for the decay in stage II as compared to exponential scaling during stage III, see App.~\ref{app:twoFluidModel},
\begin{align*}
	N_{\rm tot}(t)&=N_{\rm tot}(0)-\gamma_{\rm dec} T^\alpha t \dt
\end{align*}
The decay rate of stage II is suppressed relative to the single particle decay rate $\gamma_{\rm dec}$, at low temperatures. This separation of time scales defines the metastable state. The slope of the linear decay $\gamma_{\rm dec} T^\alpha$ highlights the fact that a perfect condensate at $T=0$ is a dark state that does not decay.

\subsection*{Conclusions}
We have observed metastable dynamics in the relaxation process of ultracold bosonic atoms from an excitation band to the band of lowest energy. By measuring the momentum resolved occupation of the bands via band mapping, we have observed three stages of dynamics, which are the relaxation dynamics to the metastable state, the metastable state itself and the relaxation to the thermal equilibrium state. Utilizing both numerical and analytical reasoning, we have provided an interpretation of this dynamical process. In particular, our simulations suggest that the metastable state is stabilized by the chiral, transient order that emerges in the excitation band. This chiral order suppresses the relaxation processes via destructive interference.

We emphasize that the mechanism described here is not only conceptually interesting, but provides a framework to create non-equilibrium ordered states in solids. By optically pumping electrons into higher bands, an ordered state can be long-lived if this ordered state is a dark state with regards to particle relaxation to the lower band, as we have argued in this work. We note that the transient order observed in this work can be related to the critical slowdown near renormalization group fixed points \cite{Mat:10, Mat:17}, to be discussed elsewhere.

\section*{ACKNOWLEDGMENTS}
M.N., R.E., and L.M. acknowledge support from the Deutsche Forschungsgemeinschaft (DFG) through the collaborative research center SFB 925. A.H. and M.H. acknowledge support through the DFG individual grant He2334/17-1. This work was partially supported by the Cluster of Excellence ’CUI: Advanced Imaging of Matter’ of the Deutsche Forschungsgemeinschaft (DFG) - EXC 2056 - project ID 390715994. M.N.~acknowledges support from Stiftung der Deutschen Wirtschaft. J.V.~is grateful to the National Agency for Research and Development (ANID) of Chile and its Ph.D.~scholarship program. We thank Jayson Cosme, Juliette Simonet and Klaus Sengstock for useful discussions.

\begin{appendix}

\section{EXPERIMENTAL PROTOCOLS}\label{app:Exp}
\begin{figure}[tb]
\centering
\includegraphics[width=1\linewidth]{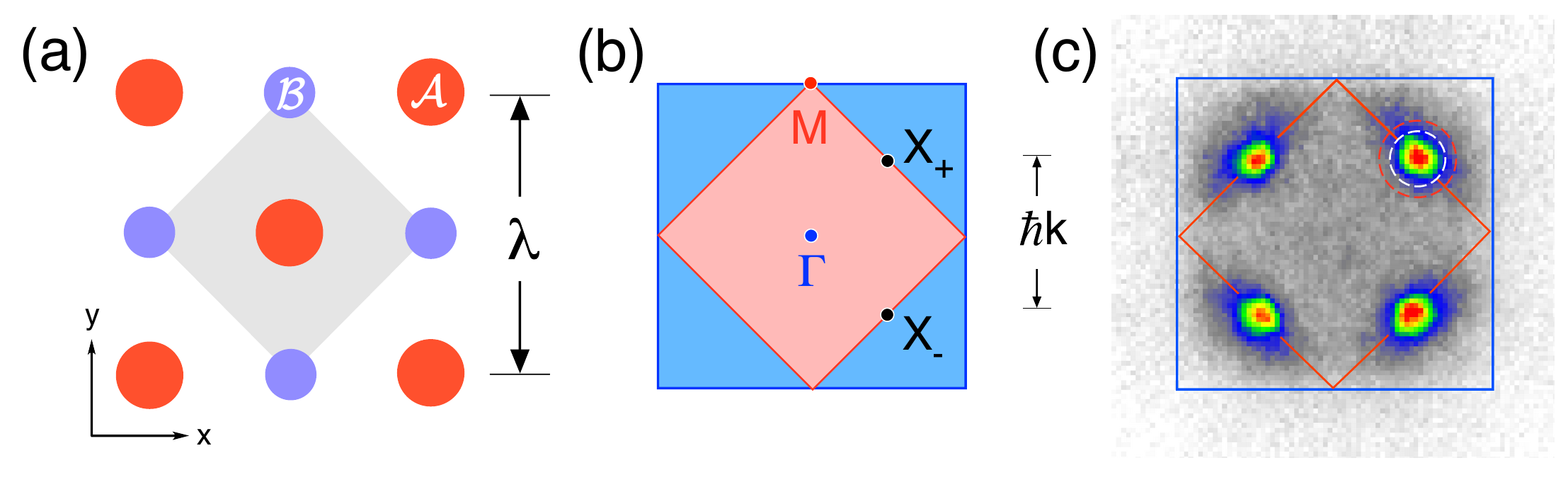}
\caption{(a) Lattice geometry and unit cell (gray area). (b) Sketch of 1st (red area) and 2nd (blue area) Brillouin zones with the high symmetry points $\Gamma$, M, $X_{+}$, and $X_{-}$. (b) Sketch of the disk-shaped (area within dashed white circle) and ring-shaped (area within red and white dashed circles) regions of interest used for data evaluation.}
\label{fig:sketchBZ}
\end{figure}

The light-shift potential for the bipartite square optical lattice reads
\begin{align}
&V(x,y)=-V_0 \bigg[\cos^2(k x) + \cos^2(k y) \eqb
+ 2 \cos(\theta) \cos(k x)\cos(k y) \bigg] \label{eq:LatticePotential} \com
\end{align}
where $V_0 (1 + 2\cos(\theta))$ is the total depth of the lattice and $k=2\pi/\lambda$ with wavelength $\lambda = 1064\,$nm, see Ref.~\cite{Koc:16}. The lattice is composed of two classes of wells, denoted $\mathcal{A}$ and $\mathcal{B}$, with different depth arranged according to the black and white fields of a chequerboard, see Fig.~\ref{fig:sketchBZ}(a). The control parameter $\theta$, that can be experimentally adjusted on a few-ten-microsecond time-scale with better than $\pi/300$ precision by interferometric techniques \cite{Hem:91}, determines the difference of the depths of the $\mathcal{A}$- and $\mathcal{B}$-sublattice sites, which is given by $\Delta V = -4V_0 \cos(\theta)$. In its second Bloch band this lattice provides two inequivalent degenerate energy minima, denoted $X_{+}$ and $X_{-}$, at the edge of the first Brillouin zone (BZ), see Fig.~\ref{fig:lattice}(b) in the main text and Fig.~\ref{fig:sketchBZ}(b). 

A Bose-Einstein Condensate (BEC) of $10^5$ $^{87}{\rm Rb}$ atoms in the $|{\rm F}=2, {\rm m_{F}}=2>$ hyperfine state is prepared in an isotropic magnetic trap with trap frequencies ($\omega_{x}$,$\omega_{y}$,$\omega_{z}$)$=(39,42,35)\,{\rm Hz}$. By adiabatically ramping up the lattice depth in $100{\,\rm ms}$ to $V_{0}=7.2\,{\rm E_{rec}}$, the atoms are loaded into the lowest band of the optical lattice. The time phase at this stage is set to $\theta=0.4\,\pi$ corresponding to a potential difference $\Delta V = \usi{-1.2}{V_0}$. After a short waiting time of $10\,{\rm ms}$, $\theta$ is tuned in $300\,{\rm \mu s}$ to a final value 
 via a band-mapping technique: The lattice potential is adiabatically ramped down to zero in $2\,{\rm ms}$ followed by a ballistic expansion during $30\,{\rm ms}$. This maps the quasimomenta of the Bloch functions onto the momenta of the free particles, thus giving rise to a map of the populations of the Brillouin zones (BZ) of the lattice, which is recorded via an absorption image.

To determine the condensed part of the atomic sample, we define four disk-shaped regions of interest (ROIs) and four ring-shaped ROIs, see Fig.~\ref{fig:sketchBZ}(c), all enclosing the same area, centered around the $X_{\pm}$-points. By subtracting the number of atoms found within the outer ring from the atoms within the inner disk, we get rid of the thermal part and remain with the number of condensed atoms. The number of thermal atoms in the first and second band are determined by counting the atoms within ROIs comprising the first and second BZs, respectively, however, with the intersections with the disk-shaped ROIs around $X_{\pm}$ cut out and with the intersections with the ring-shaped ROIs counted twice. For the total population in the second band we add up the population of the thermal atoms in the second BZ and the condensed atoms.

For extracting the decay time scale shown in Fig.~\ref{fig:decayOfTheta}, we perform an exponential fit to the data points in stage III. We identify the onset of stage III as the time where the condensate fraction, i.e.~the number of condensed atoms divided by the number of total atoms, drops below a certain threshold value. This threshold is chosen to be 20\% for the experimental data and 5\% for the simulated data shown in Fig.~\ref{fig:decayOfTheta}. Due to collisions during the time-of-flight images and other broadening effects we expect the experimental data to overestimate the actual condensate fraction and therefore use the higher threshold value. We fit an exponential of the type $f(t)=a\,e^{-t/T_{\rm d}}+b$ and extract the decay time scale $T_{\rm d}$. 

\section{DETAILS ON CLASSICAL-FIELD-THEORY SIMULATIONS}\label{app:classicalFieldTheory}

\begin{figure*}[tb]
\centering
\includegraphics[width=0.8\linewidth]{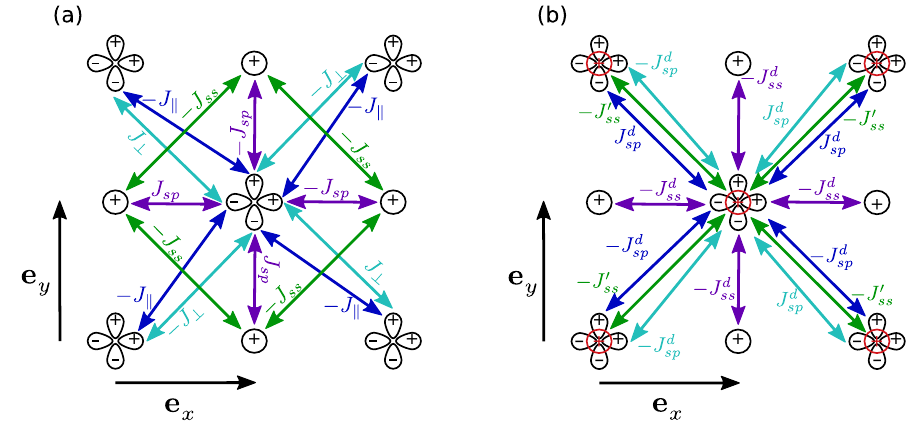}
\caption{Sketch of all hopping terms included in the Hamiltonian $H_{\rm xy}$ for each $xy$-plane. Circles denote $s$-orbitals while dumbbells denote $p$-orbitals. Next to each hopping strength we indicate the two tight-binding orbitals involved in the corresponding hopping process by an arrow of the same color. Panel (a) shows all hopping terms between $s$-orbitals on $\mathcal A$ sublattice sites and $p$-orbitals on $\mathcal B$ sublattice sites, while panel (b) shows all hopping terms that involve at least one $s$-orbital on a $\mathcal B$ sublattice site. The latter $s$-orbitals are indicated by red circles in panel (b).}
\label{fig:sketchHopping}
\end{figure*}

For our numerical simulations we employ a tight-binding model that includes both the dynamics between individual tubes within the $xy$-plane as well as the dynamics within each tube in the $z$-direction. We first describe how we model the dynamics within a single $xy$-plane and afterwards explain how we extend this to include the dynamics in the tubes along the $z$-direction.

For the dynamics within the $xy$-plane we use a lattice with $12\!\times\!12$ sites with periodic boundary conditions. In contrast to the experimental setup we do not include an additional weak harmonic trap for our simulations. Instead we only model the dynamics in the center of the trap where the additional trapping potential is flat to good approximation. In the Hamiltonian $H_{\rm xy}$ for each $xy$-plane we include all nearest- and next-nearest-neighbour hopping terms as indicated in Fig.~\ref{fig:sketchHopping}. Additionally we include on-site interaction terms. After the quench that excites the atoms to higher bands, such that $\Delta V\gg 0$, the lowest band is almost entirely composed of $s$-orbitals on $\mathcal B$ sublattice sites. Keeping this in mind, it is convenient to group the terms of $H_{\rm xy}$ into four subsets
\begin{align}
	H_{\rm xy}&=H_{\rm lb} + H_{\rm eb} + H_{\rm nn} + H_{\rm exc}
\end{align}
The first subset $H_{\rm lb}$ denotes those terms that only involve $s$-orbitals on $\mathcal B$ sublattice sites, hence those orbitals that comprise the lowest band for $\Delta V>0$. Its explicit form is
\begin{align}
	H_{\rm lb}&=-J_{ss}'\hspace{-10pt} \sum_{\substack{\mbf R \in \mathcal B,\\ \mbf d=\pm \mbf e_x',\pm \mbf e_y'}} \hspace{-10pt} b_{\mbf R+\mbf d,s}^\dag b_{\mbf R,s} + V_s' \sum_{\mbf R \in \mathcal B} b_{\mbf R,s}^\dag b_{\mbf R,s} \eqb
	+\frac{U_B}{2} \sum_{\mbf R\in \mathcal B} b_{\mbf R,s}^\dag b_{\mbf R,s}^\dag b_{\mbf R,s} b_{\mbf R,s}\dt
\end{align}
Here the operator $b_{s,x,y,\mbf R}^\dag$ ($b_{s,x,y,\mbf R}$) creates (annihilates) an atom on the site $\mbf R$ in the $s$- $p_x$- or $p_y$-orbital, respectively. 
The unit vectors $\mbf e_x$ and $\mbf e_y$ connect $\mathcal{A}$ and $\mathcal{B}$ sites along the $x$ and $y$ directions, respectively, $\mbf e_x'=\mbf e_x-\mbf e_y$ and $\mbf e_y'=\mbf e_x+\mbf e_y$. The term $H_{\rm lb}$ involves a hopping term between nearest neighbour $\mathcal B$-sites with strength $J_{ss}'$, an on-site potential $V_s$ and an interaction term.

The second subset $H_{\rm eb}$ involves only those orbitals that comprise the excited bands
\begin{widetext}
\begin{align}
H_{\rm eb}&=-J_{ss} \sum_{\substack{\mbf R \in \mathcal A,\\ \mbf d=\pm \mbf e_x',\pm \mbf e_y'}} b_{\mbf R+\mbf d,s}^\dag b_{\mbf R,s} + V_s \sum_{\mbf R \in \mathcal A} b_{\mbf R,s}^\dag b_{\mbf R,s}\eqb
+J_{sp}\sum_{\mbf R \in \mathcal B} - b_{\mbf R+\mbf e_x,s}^\dag b_{\mbf R,x} + b_{\mbf R-\mbf e_x,s}^\dag b_{\mbf R,x} - b_{\mbf R+\mbf e_y,s}^\dag b_{\mbf R,y} + b_{\mbf R-\mbf e_y,s}^\dag b_{\mbf R,y}+ {\rm h.c.} \eqb
- J_{\parallel}\sum_{\substack{\mbf R \in \mathcal B,\nu\in\{x,y\}\\\mbf d =\pm \mbf e_x',\pm \mbf e_y'}} b_{\mbf R+\mbf d,\nu}^\dag b_{\mbf R,\nu} + \sum_{\nu} V_\nu \sum_{\mbf R \in \mathcal B} b_{\mbf R,\nu}^\dag b_{\mbf R,\nu} 
-J_{\perp} \sum_{\substack{\mbf R\in \mathcal B,\\ \mbf d=\pm \mbf e_x', \pm \mbf e_y'}} b_{\mbf R+\mbf d,x}^\dag b_{\mbf R,y} + {\rm h.c.}\eqb
\frac{U_A}{2} \sum_{\mbf R\in \mathcal A} b_{\mbf R,s}^\dag b_{\mbf R,s}^\dag b_{\mbf R,s} b_{\mbf R,s} + \frac{3 U_B}{8} \sum_{\substack{\mbf R \in \mathcal B,\\\nu\in\{x,y\}}}  b_{\mbf R,\nu}^\dag b_{\mbf R,\nu}^\dag b_{\mbf R,\nu} b_{\mbf R,\nu} 
+\frac{U_B}{2} \sum_{\mbf R \in \mathcal B}  b_{\mbf R,x}^\dag  b_{\mbf R,x} b_{\mbf R,y}^\dag b_{\mbf R,y}\eqb
+\frac{U_B}{8} \sum_{\mbf R \in \mathcal B}\vsb{ b_{\mbf R,x}^\dag b_{\mbf R,x}^\dag b_{\mbf R,y} b_{\mbf R,y} +  {\rm h.c.} }\dt
\end{align}
\end{widetext}
Here all combinations of hopping processes between $s$-orbitals on $\mathcal A$-sites and $p$-porbitals on $\mathcal B$-sites as well as interaction and on-site potentials for these orbitals are included.
The third subset $H_{\rm nn}$ contains only one density-density-type interaction term that does involve both the lowest and excited bands, but leaves each of their occupations unchanged
\begin{align}
H_{\rm nn}&=U_B \sum_{\substack{\mbf R \in \mathcal B,\\\nu\in\{x,y\}}} b_{\mbf R,\nu}^\dag b_{\mbf R,\nu} b_{\mbf R,s}^\dag b_{\mbf R,s}\dt
\end{align}
The fourth subset $H_{\rm exc}$ contains the most interesting terms, as these lead to an exchange of particles between excited bands and the lowest band. Hence only these terms can lead to decay from excited bands to the lowest band. The explicit form is
\begin{align}
H_{\rm exc}&= J^d_{sp} \sum_{\substack{\mbf R \in \mathcal B,\\\nu\in\{x,y\}}}  \vb{b_{\mbf R-\mbf e_\nu',s}^\dag-b_{\mbf R+\mbf e_\nu',s}^\dag} b_{\mbf R,x} + {\rm h.c.}\eqb
+ J^d_{sp} \sum_{\substack{\mbf R \in \mathcal B,\\\sigma=\pm 1}}  \vb{\sigma b_{\mbf R+ \sigma \mbf e_x',s}^\dag+ \sigma b_{\mbf R-\sigma \mbf e_y',s}^\dag} b_{\mbf R,y} + {\rm h.c. }\eqb
-J_{\rm ss}^d W_{\rm hop}
+ \frac{U_B}{4} W_{\rm int} + {\rm h.c.}
\end{align}
We find numerically that the first two terms proportional to $J_{sp}^d$ are negligible for $\Delta V>0$. Hence the only two terms that can lead to decay from excited bands to the lowest band are the latter two terms with the operators
\begin{align}
W_{\textrm{int}} &\equiv   \sum_{\mbf R \in \mathcal{B}} b_{{\mbf R},s}^\dag b_{{\mbf R},s}^\dag \vb{b_{{\mbf R},x}b_{{\mbf R},x}+b_{{\mbf R},y} b_{{\mbf R},y}} \\
W_{\textrm{hop}} &\equiv  \sum_{\substack{\mbf R \in \mathcal{B},\\ \nu \in \{x,y\}}} b_{\mbf R,s}^\dag  (b_{{\mbf R}-{\mbf e}_{\nu} ,s} + b_{{\mbf R}+{\mbf e}_{\nu},s}  )
\end{align}
as also defined in the main text. For all cases the relative prefactors of the interaction terms have been obtained by computing the overlap of harmonic oscillator wave functions for $s$- and $p$-orbitals in the $xy$-plane.

\begin{figure*}[tb]
\centering
\includegraphics[width=0.8\linewidth]{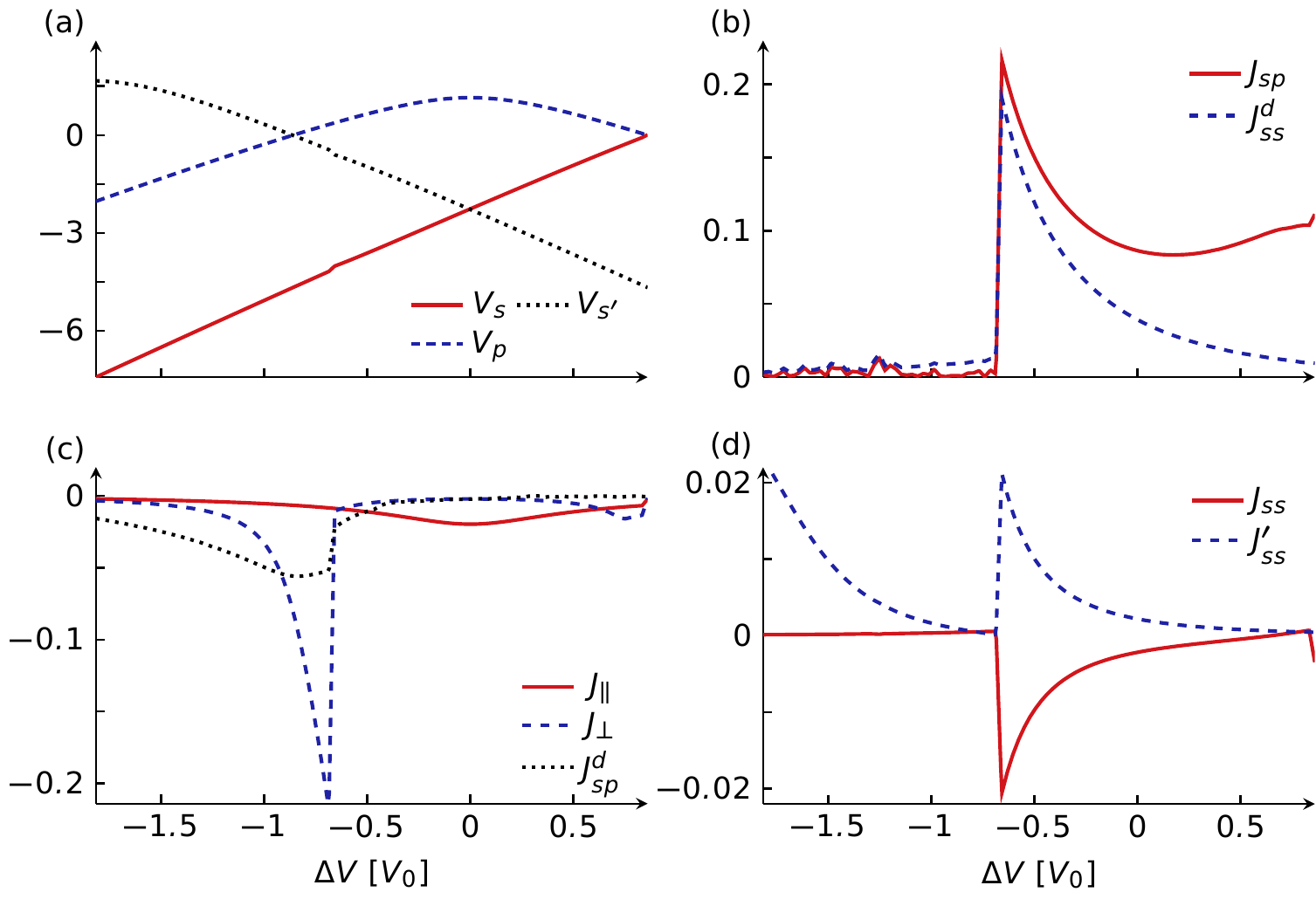}
\caption{Tight-binding parameters as a function of $\theta$ for $V_0=7 E_{\rm rec}$, $\lambda_L=\usi{1064}{\nano\meter}$ and the rubidium mass $m=87m_p$. All parameters are given in units of $E_{\rm rec}$.}
\label{fig:tightBindingParams}
\end{figure*}

Next we describe how we adjust the tight-binding parameters. For each value of $\Delta V$ we adjust all hopping terms and on-site potentials such that the resulting tight-binding band structure has optimal agreement with the corresponding numerically determined Bloch band structure at certain symmetry points. For our case we use the symmetry points $\Gamma$, $M$, $X$ and the points half way between $\Gamma$ and $M$, as indicated in Fig.~\ref{fig:sketchBZ}(b). Some of the tight-binding parameters we determine from analytical expressions for the tight-binding band structure at these symmetry points. Others we determine numerically using a variational Monte-Carlo-type optimization routine. The resulting set of hopping parameters and on-site potentials for $V_0=7\, E_{\rm rec}$ is shown in Fig.~\ref{fig:tightBindingParams}. For the interaction parameters we determine the effective one-dimensional interaction strength of each tube as $g_{1d,C}=2\hbar^2 a/(mx_{0,C}^2)$, see e.g. \cite{Pet:08}, where $a=5\,$nm is the three-dimensional scattering length and $m$ the mass of $^{87}$Rb, $C=\mathcal A, \mathcal B$ and $x_{0,C}$ is the harmonic oscillator length. We determine the harmonic oscillator length by expanding the lattice potential around each well and comparing to a harmonic oscillator potential. Depending on the value of $\Delta V$ and the sublattice site considered we obtain values in the range $\usi{0.04}{\micro\meter}\,E_{\rm rec}<g_{1d}<\usi{0.06}{\micro\meter}\,E_{\rm rec}$.

\begin{figure*}[tb]
\centering
\includegraphics[width=0.9\linewidth]{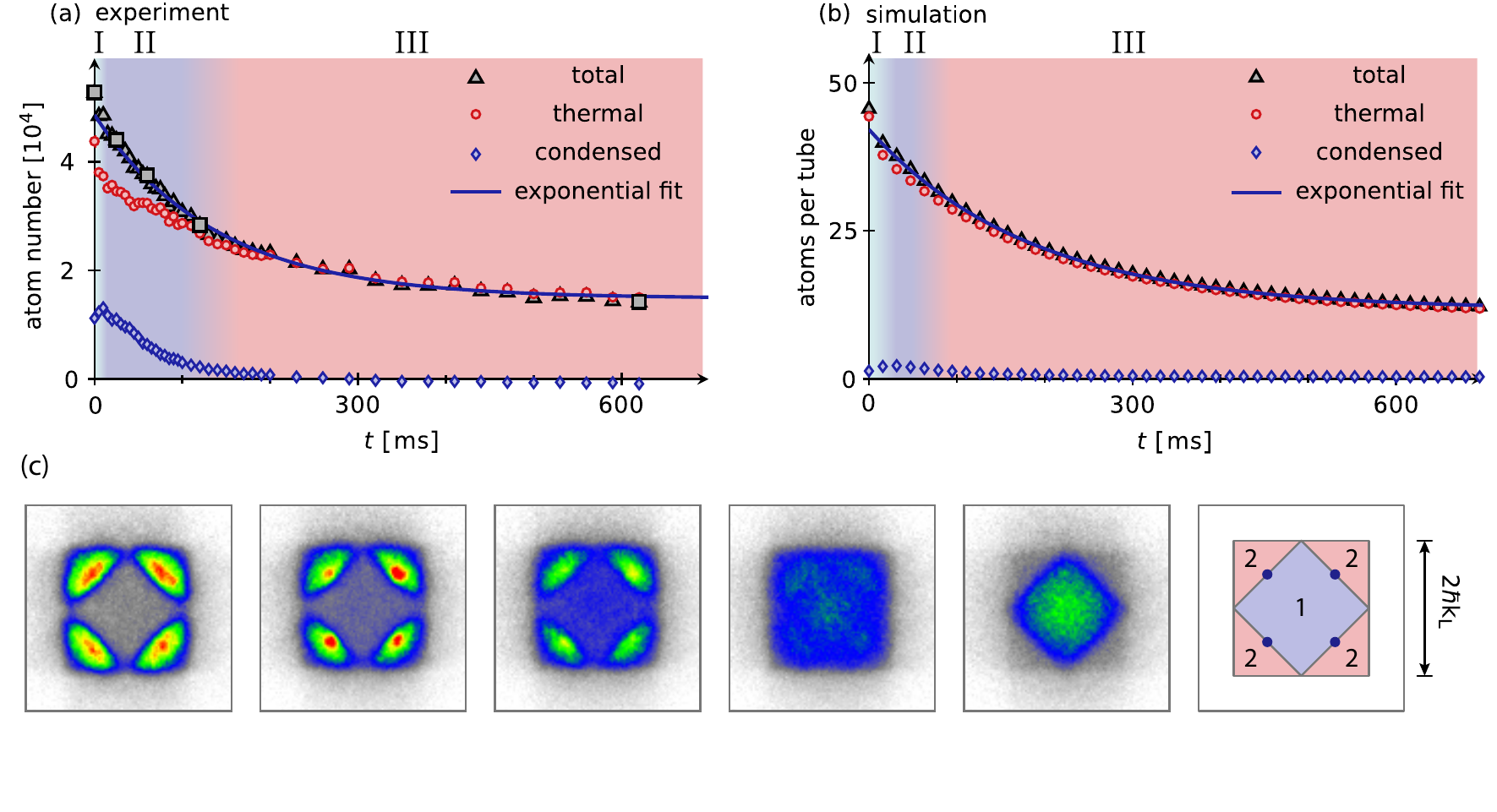}
\caption{Time evolution of the total number of atoms in the second band as well as its condensate and thermal fraction after a quench to $\Delta V=\usi{0.43}{V_0}$. Panel (a) shows experimental and panel (b) numerical data. We identify three main decay stages: the coherence buildup stage (I), the inhibited relaxation stage (II) and the fast relaxation stage (III). The blue lines show exponential fits to the data points in stage III. (c) Time of flight images for experimental data points marked by black squares in (a). The temperature of the initial state for both experiments and simulation is $T\approx\usi{1.1}{E_{\rm rec}/k_B}\approx\usi{110}{\nano\kelvin}$. We note that atom numbers for experiment and simulation are adjusted to be comparable in the center of each tube as is described in Appendix \ref{app:classicalFieldTheory}.}
\label{fig:selfStabilizationHot}
\end{figure*}

In order to include the dynamics within each tube along the $z$-direction we artificially introduce a lattice with spacing $d_z$. The artificial discretization allows to impose a tight-binding model with hopping constant $J_z$ and lattice potential $V_z$ along the $z$-direction. This constitutes a reasonable approximation as long as the discretization length $d_z$ is small compared to the healing length of the condensate and the thermal de-Broglie wavelength. For our case this is fulfilled for $d_z=0.13\,\mu$m. The full Hamiltonian then splits into a part for each $xy$-plane and the part for the $z$-direction
\begin{align}
	H&=H_{\rm xy} + H_{\rm z}\\
	H_{\rm z}&=-J_z \sum_{\substack{\mbf R,\nu\in\{s,x,y\},\\\mbf d \in\{\pm \mbf e_z\}}} b_{\mbf R + \mbf d,\nu}^\dag b_{\mbf R,\nu} \eqb
	+ V_z \sum_{\mbf R,\nu\in\{s,x,y\}} b_{\mbf R,\nu}^\dag b_{\mbf R,\nu} \dt
\end{align}
By requiring that the lowest tight-binding band of the artificial lattice for small quasimomenta is approximately equal to the free particle dispersion relation, we obtain for the hopping constant $J_z=\hbar^2/(2m d_z^2)=\usi{1.7}{E_{\rm rec}}$. Since it is numerically too expensive to consider a trap in the $z$-direction that is as large as in the experiment, we instead determine the density in the center of the trap $n_{\rm cent}$ from a Thomas-Fermi approximation and adjust the occupation of the corresponding lattice  sites $n_{C}$, $C=\mathcal A, \mathcal B$, such that $n_{C}=n_{\rm cent} d_z$. Note that $n_{\rm cent}$ depends on $C$ via $g_{1d}$. For $300$ atoms per tube we find $n_C\approx 3$. Furthermore, we choose the interaction strengths $U_C$, for $C=\mathcal A, \mathcal B$, such that $U_C \, n_C=g_{1d} n_{\rm cent}$. We obtain $U_A\approx U_B\approx \usi{0.4}{E_{\rm rec}}$. We use a lattice of $N_z=25$ sites in the $z$-direction and adjust the trapping potential $V_z$ such that the occupations $n_C$ vanishes at the edge of the trap. Hence we use a trap that is shorter and steeper as compared to the experimental values, while at the same time matching the density and mean-field interaction in the center of the trap. This leads to an overall lower number of atoms per tube as can be seen in Fig.~\ref{fig:selfStabilizationHot} and in Fig.~\ref{fig:selfStabilization} in the main text.

For the time evolution we start at $\theta=0.35\,\pi$ corresponding to $\Delta V\approx\usi{-1.8}{V_0}$ and use a classical-field approach, for reviews see \cite{Bla:08, Gar:01, Gar:03, Coc:09, Pol:10, Dav:11}. For large occupation of modes it is a good approximation to replace creation and annihilation operators by their expectation values and solve the resulting Heisenberg equations of motion for these operators numerically. We initialize the system using Monte-Carlo minimization, starting from an empty lattice and working at fixed chemical potential. For the Monte-Carlo procedure we use the Hamiltonian as the minimization functional and adjust the Monte-Carlo temperature to the desired temperature of the lattice. For the low temperature data we perform annealing starting at higher temperature and reducing the temperature to the desired one during the Monte-Carlo procedure. For our simulations we repeat this procedure and the subsequent time evolution $300$--$600$ times. We average all observables over all of these Monte-Carlo trajectories. This accounts for thermal fluctuations. Note that a rather low number of Monte-Carlo trajectories is sufficient due to the self-averaging along the $z$-direction.			
We read out the occupations of individual bands by projecting onto the Bloch functions of the tight-binding lattice. As an estimate for the condensed atoms we use the number of atoms occupying the $X$-points of the lattice.

\section{TWO-FLUID MODEL}\label{app:twoFluidModel}
\begin{figure*}[tb]
\centering
\includegraphics[width=0.9\linewidth]{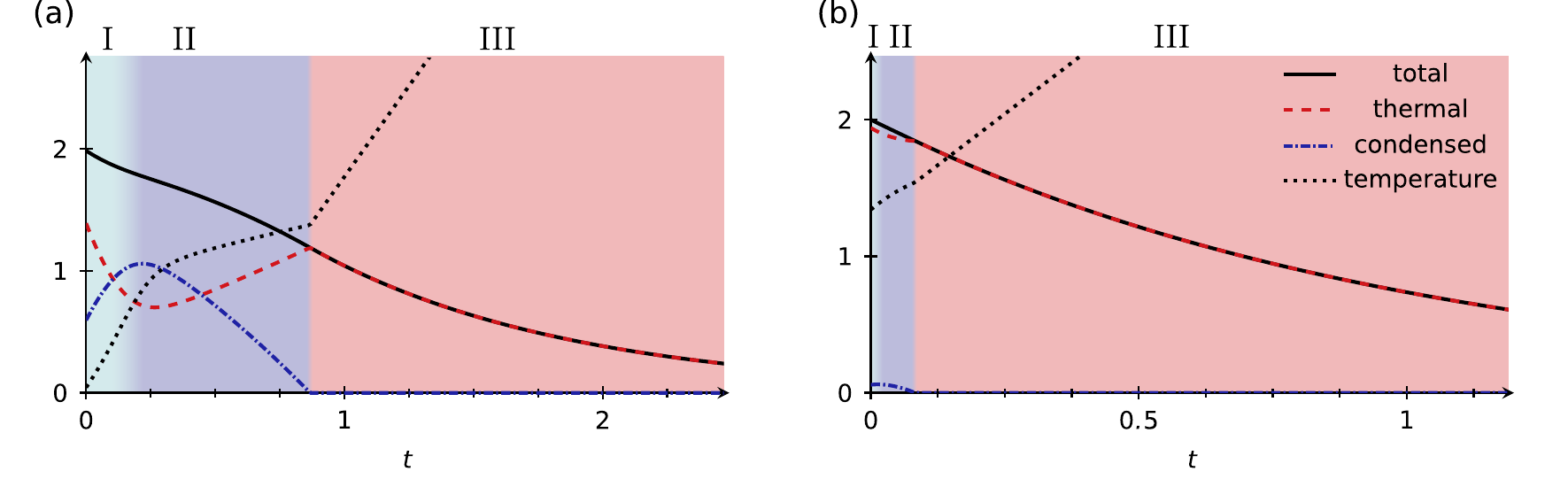}
\caption{Two-fluid model for self-stabilization at low initial temperature (a) and high initial temperature (b). We show the total, thermal and condensed particle number as well as temperature computed within a two-fluid model, for details see App.~\ref{app:twoFluidModel}. The model shows qualitative agreement with experimental and simulation data shown in Fig.~\ref{fig:selfStabilization} of the main text. All quantities within the two-fluid model are unitless. We use $\gamma_{\rm dec}=1$, $\Gamma_{\rm eq}=3$, $\alpha=2$ and $\Delta E=3$.}
\label{app:fig:twoFluidModel}
\end{figure*}
A simple two fluid model captures the main aspects of the three different decay stages and qualitatively shows the same behavior as the experimental data and our full simulation. We consider only the atoms in the upper band and assume that we have a thermal fraction $N_{\rm th}$ and a condensed fraction $N_{\rm c}$ of atoms. The total number of atoms is $N_{\rm tot}=N_{\rm th}+N_{\rm c}$. Furthermore, we assume that the equilibration within the thermal and condensed atoms happens on a much faster time scale than the equilibration between the two, such that we can assume the equilibration within each fraction to be instantaneous. Without loss of generality we assume that the condensed atoms have zero energy. Following the description in chapter 2 of Ref.~\cite{Pet:08} we assume a generic density of states
\begin{align}
g(\epsilon)&= c_\alpha \epsilon^{\alpha-1}\dt
\end{align}			
We can then compute the total number of thermal atoms and the total energy $E$ as
\begin{align}
N_{\rm th}&=\int \D \epsilon\; g(\epsilon) \frac{1}{e^{\epsilon/(k_BT)}-1}\propto T^\alpha\\
E&=\int \D \epsilon \; g(\epsilon) \frac{\epsilon}{e^{\epsilon/(k_B T)}-1}\propto T^{\alpha+1} \dt
\end{align}
We absorb the proportionality constants into the units of temperature and energy and therefore obtain
\begin{align}
N_{\rm th}&= T^\alpha & N_{\rm th } T &= E \dt
\end{align}
As we have seen above the condensed atoms do not decay due to perfect destructive interference. We therefore assume that only the thermal atoms decay with time-scale $1/\gamma_{\rm dec}$. Furthermore, we assume that the thermal and condensed fraction equilibrate on a time scale of $1/\Gamma_{\rm eq}$. On average whenever a thermal atom decays the total energy is decreased by the mean energy of a thermal atom $E/N_{\rm th}$. Additionally the atom gains an energy $\Delta E$ corresponding to the energy difference between the upper and the lower band. We assume that this energy is redistributed to the atoms in the upper band and hence the energy of these atoms is increased by this amount. Hence the equations of motion for our model system are
\begin{align}
\dot N_{\rm th}&=-\gamma_{\rm dec} N_{\rm th} + \Gamma_{\rm eq} (T^\alpha-N_{\rm th}) \nonumber \\
\dot N_{\rm tot}&=-\gamma_{\rm dec} N_{\rm th} \nonumber \\
\dot E&=(\Delta E- \frac{E}{N_{\rm th}}) \gamma_{\rm dec} N_{\rm th} \label{eq:app:de} \dt
\end{align}%

\begin{figure}[!tb]
\centering
\includegraphics[width=0.9\linewidth]{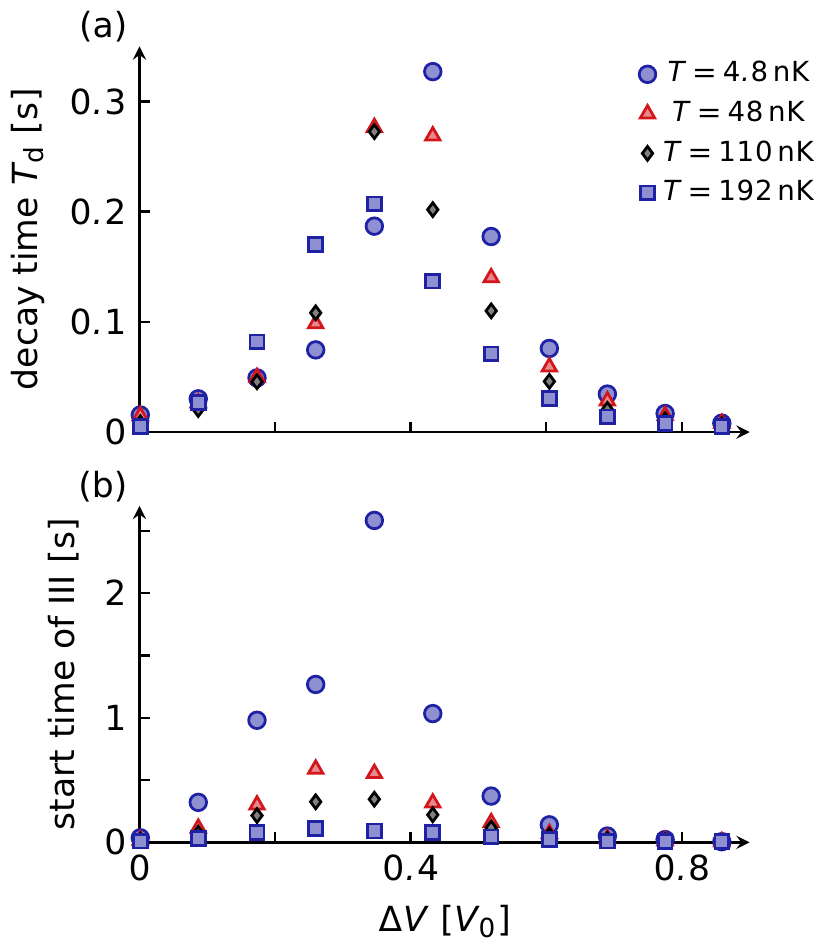}
\caption{(a) Simulated decay time scale as a function of final potential offset $\Delta V$ at several different temperatures as indicated in the legend. After loading the atoms into the second Bloch band they relax back to the lowest band. During stage III the decay is exponential and we show  the associated time scale versus $\Delta V$. (b) Cross-over time from stage II-III. We extract the time where the condensate fraction drops below 3\%, which indicates the cross-over from stage II to III and is the starting point for the exponential fits. We see that the cross-over time crucially depends on temperature while the temperature dependence of the decay time scale is not as strong.}
\label{fig:decayOfThetaDifT}
\end{figure}

We derive the equation for the temperature of the system by taking the derivative of $T=E/N_{\rm th}$. We obtain
\begin{align}
\dot T&=\gamma_{\rm dec} \Delta E - \frac{\Gamma_{\rm eq} T}{N_{\rm th}} (T^\alpha-N_{\rm th}) \label{app:eq:difT} \dt
\end{align}
These equations are of cause only valid as long as there is a condensed fraction, otherwise $N_{\rm tot}=N_{\rm th}$ and
\begin{align}
\dot N_{\rm th}&=-\gamma_{\rm dec} N_{\rm th}\\
\dot T&=\gamma_{\rm dec} \Delta E\dt
\end{align}
In our case, having one harmonically-trapped and two free dimensions, we obtain $\alpha\approx 1/2 + 1/2 + 1=2$. 

From Eq.~\ref{app:eq:difT} we can see that moving an atom from the condensate to the thermal cloud effectively cools the atoms. The reason is that the mean kinetic energy of thermal atoms is larger than the mean kinetic energy of condensed atoms. This cooling process counteracts the heating effect due to the decay of thermal atoms. 		

Within the two-fluid model we can also obtain an approximation for the scaling of the decay of total atoms. To this end we assume that the equilibration time scale $\Gamma_{\rm eq}$ is much faster than all other time scales, such that $N_{\rm th}\approx T^\alpha$. We find that this approximation is reasonable during stage II. Within this approximation we can solve Eq.~\ref{eq:app:de} and derive the corresponding time dependence of the temperature
\begin{align}
T(t)&=(\Delta E- (\Delta E-T_0) e^{-\frac{\gamma}{\alpha+1}t})^{\alpha}
\end{align}
Hence the temperature always equilibrates at $\Delta E^\alpha$. 
For any fixed temperature we can now solve the equation of motion for the total number of atoms
\begin{align}
N_{\rm tot}(t)&=N_{\rm tot}(0)-\gamma_{\rm dec} T^\alpha t \dt
\end{align}
and see that it scales linearly. We note that for large $\Delta E$ the condensate fraction may be depleted before the constant temperature is reached and hence linear scaling of the total number of atoms is not always observed.

For completeness we also show the results of the two-fluid model for lower and higher initial temperature in Fig.~\ref{app:fig:twoFluidModel}. By construction, the decay of the two-fluid model is exponential when the condensate fraction vanishes. We also see from Fig.~\ref{app:fig:twoFluidModel} that the inflection point of the curve for the total number of excited atoms is exactly the point where the condensate fraction vanishes.

\section{DECAY OF THERMAL STATE}\label{app:thermalDecay}
As a cross-check that the coherence of the chiral condensate is indeed the origin of the inhibition of decay, we consider a cloud of atoms at higher initial temperature in Fig.~\ref{fig:selfStabilizationHot}. As a result we obtain a lower condensate fraction and larger thermal fraction. The decay is now dominated by the thermal fraction of the atoms. We therefore find that an exponential fit shows good agreement even for short times. This is in contrast to the low temperature sample discussed in Fig.~\ref{fig:selfStabilization} of the main text.
\\
\section{DECAY-TIME-SCALE FOR DIFFERENT TEMPERATURES}
\label{app:decayTimeScaleDiffT}
Within our numerical simulations we can determine the decay time for the exponential decay in stage III for a range of different temperatures. The results are shown in Fig.~\ref{fig:decayOfThetaDifT}(a). The decay time shows no strong dependence on the initial temperature of the atomic cloud. Only the maximum of the decay times shifts to slightly lower values of $\Delta V$ for higher temperatures. This can be explained as follows: the initial temperature of the cloud determines the condensate fraction at the beginning of stage II. Hence we expect larger coherence and slower decay during stage II. Subsequently heating leads to increasing temperature and hence reducing phase-space density. The onset of stage III is essentially determined by the time when the phase-space density has reduced below the critical value for condensation. Hence independent of the initial temperature the phase-space density in the beginning of stage III is always the same. A lower initial temperature only leads to a later cross-over from stage II to III. We confirm this by showing the cross-over time point from stage II to III in Fig.~\ref{fig:decayOfThetaDifT}, which indeed changes dramatically with temperature, indicating significantly longer durations of stage II and hence longer life-times of the condensate for lower initial temperature. 

\section{COMPARISON OF DECAY FOR DIFFERENT TUBE LENGTHS}\label{app:compareTubeLength}
\begin{figure}[!tb]
\centering
\includegraphics[width=0.9\linewidth]{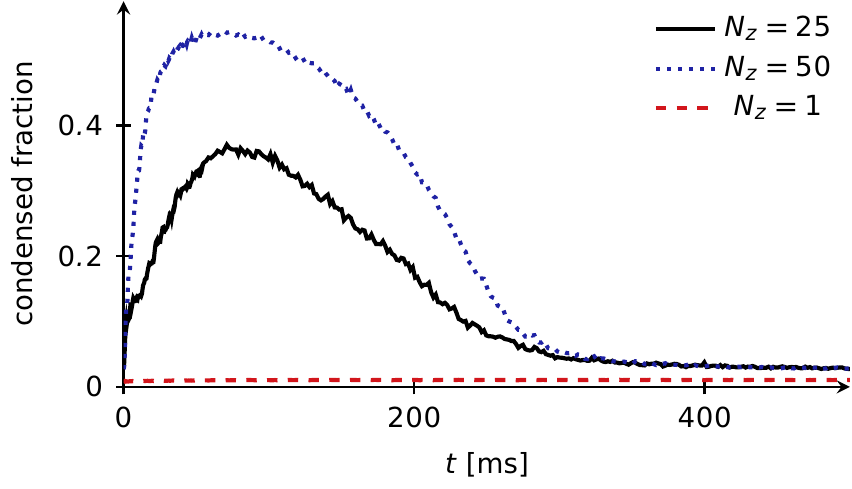}
\caption{Condensed fraction of atoms as a function of time for three different tube lengths. In order to adjust for different atom numbers per tube we plot the condensed fraction, which is the number of condensed atoms normalized with the total number of atoms in the second band. We consider three different numbers of sites in the $z$-direction $N_z$ as indicated in the legend. The corresponding length of the tubes is obtained by multiplying with the discretization length $d_z=0.13\;\mu$m, for details see App.~\ref{app:classicalFieldTheory}. Both the condensation speed and the maximal number of condensed atoms depend crucially on $N_z$. The temperature of the initial state is $T\approx\usi{0.5}{E_{\rm rec}/k_B}\approx\usi{50}{\nano\kelvin}$.}
\label{fig:app:decayOfNz}
\end{figure}
As described in App.~\ref{app:classicalFieldTheory} we use shorter tubes in the $z$-direction in our numerical simulations as compared to the experimental setup, while at the same time adjusting the density and mean-field interaction in the center of the trap. This step is necessary in order to keep the numerical effort feasible. Here we compare different tube lengths in the $z$-direction. Longer tubes, while at the same time matching the density in the center and ensuring a vanishing density at the edge of the trap imply shallower traps and an overall larger number of atoms within each tube. We see in Fig.~\ref{fig:app:decayOfNz} that longer tubes lead to a faster condensation stage and a larger final condensate fraction. This is consistent with the idea that the extra dimension acts as an entropy reservoir for the condensation within the $xy$-plane. More atoms in each tube imply a larger entropy reservoir and hence more efficient cooling towards the condensed state.

Furthermore we see that no tubes in the $z$-direction, i.e.~$N_z=1$, implying an effective 2-dimensional system, result in a vanishing condensate fraction. We note that this is not generally the case. We do obtain a small condensed fraction of atoms for an effective 2-dimensional system, when starting at a much lower temperature of 5 nK in our simulations. This temperature is, however, currently unfeasible experimentally.

\end{appendix}

\end{document}